\documentclass[aps,prx,reprint,showpacs,superscriptaddress,twocolumn]{revtex4-2}
\usepackage{amssymb}
\usepackage{mathrsfs}
\usepackage{amsfonts}
\usepackage{graphicx}
\usepackage{amsmath}
\usepackage{dcolumn}
\usepackage{color}
\usepackage{subfigure}
\usepackage{epsfig}
\usepackage{soul}
\usepackage[left]{lineno}
\usepackage[colorlinks,linkcolor=blue,citecolor=blue,hyperindex,bookmarks=false,pdfstartview=FitH]{hyperref}

\usepackage{mleftright} %to get rid of the extra space around parentheses with \left and \right: write \mleft, \mright instead

\providecommand{\U}[1]{\protect\rule{.1in}{.1in}}

\newcommand{\figpanel}[2]{\hyperref[#1]{\ref*{#1}(#2)}}

\begin{document}

% \linenumbers
% \def\linenumbersep{4pt}
% \def\linenumberfont{\normalfont\scriptsize\sffamily\color{blue}}

\title{Unconventional and robust light-matter interactions \\ based on the non-Hermitian skin effect}

\author{Lei Du}
\email{lei.du@chalmers.se}
\author{Anton Frisk Kockum}
\email{anton.frisk.kockum@chalmers.se}
\affiliation{Department of Microtechnology and Nanoscience (MC2), Chalmers University of Technology, 412 96 Gothenburg, Sweden}

\date{\today}

\begin{abstract}

Lattice models featuring the non-Hermitian skin effect have attracted rapidly growing interest due to their nontrivial spectral topology and the exotic field dynamics they enable. Such non-Hermitian lattices provide a promising paradigm for engineering exotic light-matter interactions which benefit from the intrinsic chirality and unconventional (non-Bloch) band theory. Here we study a series of unconventional light-matter interactions between quantum emitters and the prototypical Hatano--Nelson model, and briefly discuss the case with an extended lattice model dubbed the bosonic Kitaev chain. We focus on the robustness of the dynamics against various imperfections and elucidate the underlying mechanisms. We consider both small emitters, which interact with the lattice at single sites, and giant emitters coupling at multiple sites. The latter exhibit an exclusive amplification mechanism, which we find enables decoherence-free dynamics even in the presence of extra dissipation in the system. The protection from dissipation arises from the cooperation of the non-Hermiticity and the self-interference effect, and is therefore lacking for small emitters. These results not only provide deeper insights into the interplay of non-Hermiticity and various interference effects, but also have potential applications in engineering exotic spin Hamiltonians and quantum networks.

\end{abstract}

\maketitle

%%%%%%%%%%%%%%%%%%%%%%%%%%%%%%%%%%%%%%%%%%%%%%%%%%%%%%

\section{Introduction}
\label{secIntro}    

Photonic lattices provide a promising platform for engineering the electromagnetic environments of quantum emitters~\cite{Lodahl2015RMP, Roy2017RMP, chiralZollerNature}, enabling the simulation of exotic spin Hamiltonians and quantum many-body systems~\cite{MB2015Douglas, MB2017Manzoni, MB2017Noh, MB2018Chang, MB2018Shi, MB2023Sheremet}. For instance, photonic lattices enable atom-photon bound states when the atomic frequencies lie within the band gaps, providing a mechanism to mediate protected interactions between quantum emitters~\cite{BS1990John, BS1994John, BS2016Calajo, BS2016Shi, BS2019Bello, BS2021Leonforte, BS2022Simone}. Moreover, photonic lattices with nontrivial topological features can serve as structured baths which endow quantum emitters with unique properties, including chiral atom-photon bound states with topological protections~\cite{BS2019Bello, Topo2021Vega, Topo2021Kim} and exotic classes of quantum entanglement between emitters~\cite{Topo2018Barik, Topo2020GE, Topo2023Vega}.

Moreover, photonic lattices can be engineered to feature various non-Hermitian band structures, as gain and loss are ubiquitous and controllable in photonic systems~\cite{NH2008Makris, NH2017Feng, NH2018Longhi, NH2018Takata, NHtutorial}. These systems can be described by non-Hermitian Hamiltonians, which give rise to a variety of peculiar phenomena, such as the coalescence of both the eigenvalues and the corresponding eigenstates at exceptional points~\cite{EP2019Miri,RyuCP2024}, anomalous topological features governed by the non-Bloch band theory~\cite{NB2016Lee, NB1, NB2, NB3, NB4, NB5}, and the non-Hermitian skin effect~\cite{Skin2018Yao, Luis2018prb, SkinTopoOrigin, YFChen2020, HigherOrderSkin,GuoCP2023,chengCP2024}. This thus opens a new frontier of quantum optics (and beyond) along the idea that shaping the field structure can lead to novel quantum optical paradigms with promising applications for quantum technologies. 

On the one hand, recent progress has demonstrated that the seemingly detrimental (structured) losses in photonic lattices can sometimes play a constructive role and even lead to counterintuitive optical phenomena~\cite{FedericoOptica,ChengLPR2024}. On the other hand, lattices featuring the non-Hermitian skin effect, such as the Hatano--Nelson (HN) model~\cite{HNmodel} and its dissipative version, do not obey the conventional Bloch theorem and thus affect the emitter dynamics in a way with no Hermitian analogue~\cite{LonghiQD, GongHN, GongHN2}. Considering the fact that the HN model features both exceptional points and the non-Hermitian skin effect~\cite{GongPRX, EpTopo}, it appears to be an excellent candidate for engineering unconventional light-matter interactions.

In particular, giant emitters (atoms)~\cite{fiveyear, LambAFK, SAW2014, NoriGA, braided}, which feature multiple separate coupling points with the bath, exhibit properties that are closely related to the dispersion relation of the field and the separations between the coupling points~\cite{ZhaoWbound, AGT2D1, AFKstructured, LeiQST, AFK2D,CYT2023prr,Leonforte2024}. Indeed, non-Hermitian lattices can endow giant emitters with even more peculiar properties~\cite{DLHN}. For example, two carefully arranged giant emitters (i.e., with an appropriate braided coupling structure) can exhibit a decoherence-free exchange interaction when they are coupled to a Hermitian lattice~\cite{AFKstructured, AFK2D}. This decoherence-free interaction, however, becomes nonreciprocal in the non-Hermitian case, with the non-reciprocity (chirality) exactly determined by the non-Hermiticity of the lattice~\cite{DLHN}. Such chiral features are essentially distinct from those induced by the additional phase difference between different coupling points of a giant emitter~\cite{WXchiral2,DLprl,CYTcp,ChiralGAprx,doublon2024}, as will be elucidated in this paper. 

Despite these intriguing phenomena, one may wonder whether these results are robust enough against system disorders and imperfections, and whether giant emitters can outperform small ones in this context. This concern arises from the fact that non-Hermitian systems can be highly sensitive to external perturbations~\cite{Aash2020nc, NHsensor2020, HNsensing2014Jan} and, in particular, the unconventional giant-emitter effects associated with the HN model are closely related to the carefully matched emitter-lattice coupling strengths~\cite{DLHN}. 

In this paper, we study the unconventional light-matter interactions between quantum emitters (both small and giant) and an HN model. We reveal the mechanisms behind a series of exotic dynamics that are related to the non-Hermiticity of the bath and discuss the robustness of the results. We find, strikingly, that the decoherence-free dynamics of giant emitters are robust against several types of disorders and perturbations, and can be preserved even in the presence of intrinsic dissipation of the lattice and the emitters themselves, implying a better \emph{dynamical protection} than in the Hermitian case. Moreover, we discuss the extremely fragile dynamics of quantum emitters coupled to a bosonic Kitaev chain~\cite{Aash2018prx, Aash2020nc, Clara2024nature, Chris2023arxiv}, which can be mapped to two independent and opposite HN models in the ideal limit. The results in this paper provide deeper insights into the interplay among non-Hermitian band structures, self and collective interferences, and the non-Hermitian skin effect, paving the way for developing novel paradigms of quantum technologies based on non-Hermitian physics.        

\begin{figure}[ptb]
\centering
\includegraphics[width=0.90\linewidth]{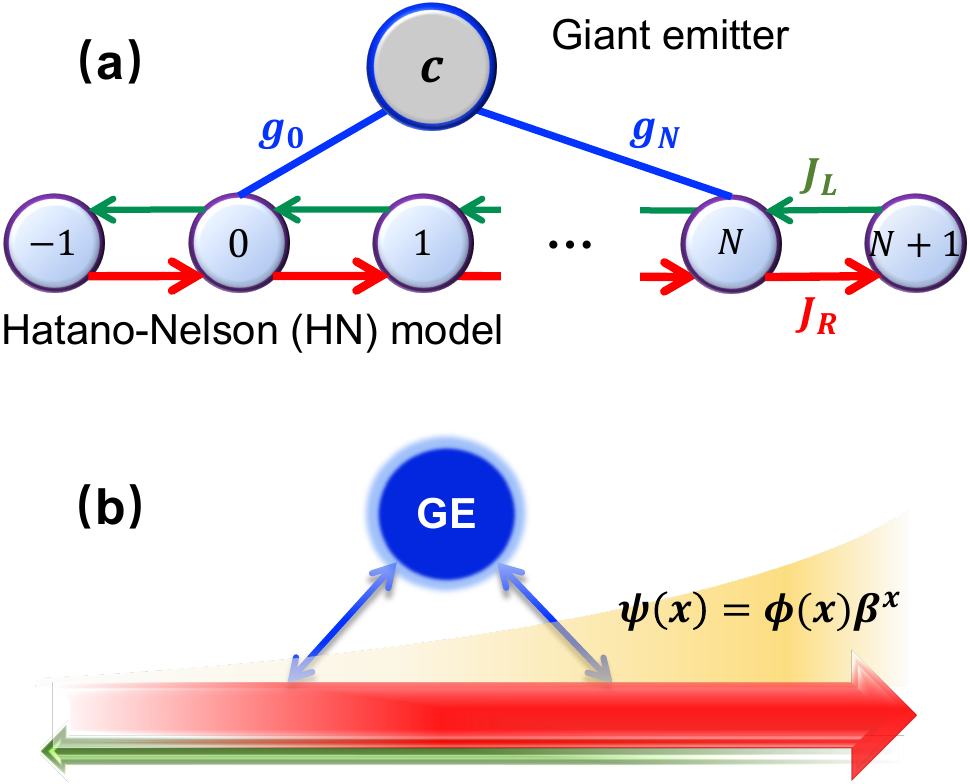}
\caption{(a) Schematic diagram of a (giant) quantum emitter $c$ coupled to a Hatano-Nelson (HN) model with asymmetric hopping rates $J_{R}$ and $J_{L}$. (b) The field coupled to the quantum emitter is equivalently subject to an imaginary gauge potential and is thus amplified along the direction of the larger hopping rate. The amplification is determined by the degree of non-Hermiticity $\beta=\sqrt{J_{R}/J_{L}}$. This corresponds to non-Hermitian skin effect under open boundary conditions.}\label{Model}
\end{figure}

%%%%%%%%%%%%%%%%%%%%%%%%%%%%%%%%%%%%%%%%%%%%%%%%%%%%%%
\section{Results and discussion}

\subsection{Basic model}
\label{secBasic}

We first consider a standard HN model, which can be described by the one-dimensional (1D) tight-binding Hamiltonian ($\hbar=1$ in this paper)
\begin{equation}
H_{\text{HN}}=\sum_{j}\mleft(J_{R}a_{j+1}^{\dag}a_{j}+J_{L}a_{j}^{\dag}a_{j+1}\mright). \label{HNH}
\end{equation}
Here $a_{j}$ ($a_{j}^{\dag}$) is the bosonic annihilation (creation) operator of the $j$th lattice site; $J_{R}=J+\gamma$ and $J_{L}=J-\gamma$ (without loss of generality, we assume both $J$ and $\gamma$ to be real) are the rightward and leftward hopping rates, respectively, which account for the nontrivial non-Hermitian topology of this model. In experiments, this type of non-Hermitian lattices have been implemented based on real-space photonic~\cite{2022laserLight} and acoustic~\cite{2021acousticNC} systems, electric ciruits~\cite{2021electricResearch,2023electricPRR}, discrete-time quantum walks~\cite{2020walkScience,2022walkNature}, and photonic synthetic dimensions~\cite{syntheticHN,LuSkin2D}. Some other promising implementation platforms, such as optomechanical~\cite{Clara2020NC,Clara2024nature} and coupled-resonator~\cite{Longhi2015prb,LonghiSR,Array2022Zhu} arrays, have also been suggested. 

One of the most important hallmarks of this non-Hermitian lattice model is the \emph{non-Hermitian skin effect} under open boundary conditions~\cite{GongPRX,EpTopo}, i.e., all the eigenstates are squeezed toward either open end of the lattice, depending on whether $J_{R}$ or $J_{L}$ is larger. More specifically, the real-space eigenstates of the HN model can be expressed as $\psi_{n}(j)=\phi_{n}(j)\beta^{j}$, where $\phi_{n}(j)$ is the usual extended Bloch wave function as in the Hermitian case (i.e., $J_{R}=J_{L}$) and $\beta=\sqrt{J_{R}/J_{L}}$ describes the degree of the non-Hermiticity. It is clear that the field is amplified, with the amplitude scaling as $\beta^{j}$, along the direction of the larger hopping rate. As long as $J_{R}$ and $J_{L}$ have the same sign, e.g., $J>\gamma>0$, a finite HN model is dynamically stable~\cite{LonghiQD,Aash2018prx,Clara2020NC} and can be mapped to a 1D pseudo-Hermitian lattice model described by the Hamiltonian
\begin{equation}
H_{\text{HN}}'=\sqrt{J_{R}J_{L}}\sum_{j}\mleft( \tilde{a}_{j+1}^{\dag}\tilde{a}_{j}+\tilde{a}_{j}^{\dag}\tilde{a}_{j+1}\mright),
\label{mapH}
\end{equation}
where the effective hopping becomes symmetric by performing an imaginary gauge transform $\tilde{a}_{n}=a_{n}\beta^{n}$~\cite{LonghiQD}. When $J_{R}$ and $J_{L}$ have opposite signs, however, the HN model becomes \emph{absolutely unstable} such that any quantum emitters coupled to this lattice, no matter how weak the coupling strengths are, always show secular energy growth~\cite{LonghiQD,DLHN}.   

Now we introduce a giant quantum emitter, which is able to interact with more than one lattice site of the HN model, as illustrated in Fig.~\ref{Model}. For concreteness, we consider a quantum harmonic oscillator (described by a bosonic mode $c$) rather than a two-level system, such that the dynamics are still linear when the emitter displays energy growth (i.e., amplification)~\cite{DLHN}. If we assume that the emitter is coupled to the sites $j=0$ and $j=N$ of the HN model with coupling strengths $g_{0}$ and $g_{N}$, respectively, the total Hamiltonian of the system is given by $H_{\text{tot}}=H_{c}+H_{\text{HN}}+H_{\text{int}}$, where
\begin{eqnarray}
H_{c}&=&\Delta_{c}c^{\dag}c, \label{Hc}\\
H_{\text{int}}&=&\mleft(g_{0}a_{0}+g_{N}a_{N}\mright)c^{\dag}+\text{H.c.} ,\label{Hint}
\end{eqnarray}
with $\Delta_{c}$ the detuning between emitter $c$ and the energy-band center of the bare HN model (assumed to be zero for simplicity). One can easily access the small-emitter case by, e.g., assuming $g_{0}\neq0$ and $g_{N}=0$, and compare the resulting behaviors with those in the giant-emitter case of $\{g_{0},\,g_{N}\}\neq0$. 

\begin{figure}[ptb]
\centering
\includegraphics[width=\linewidth]{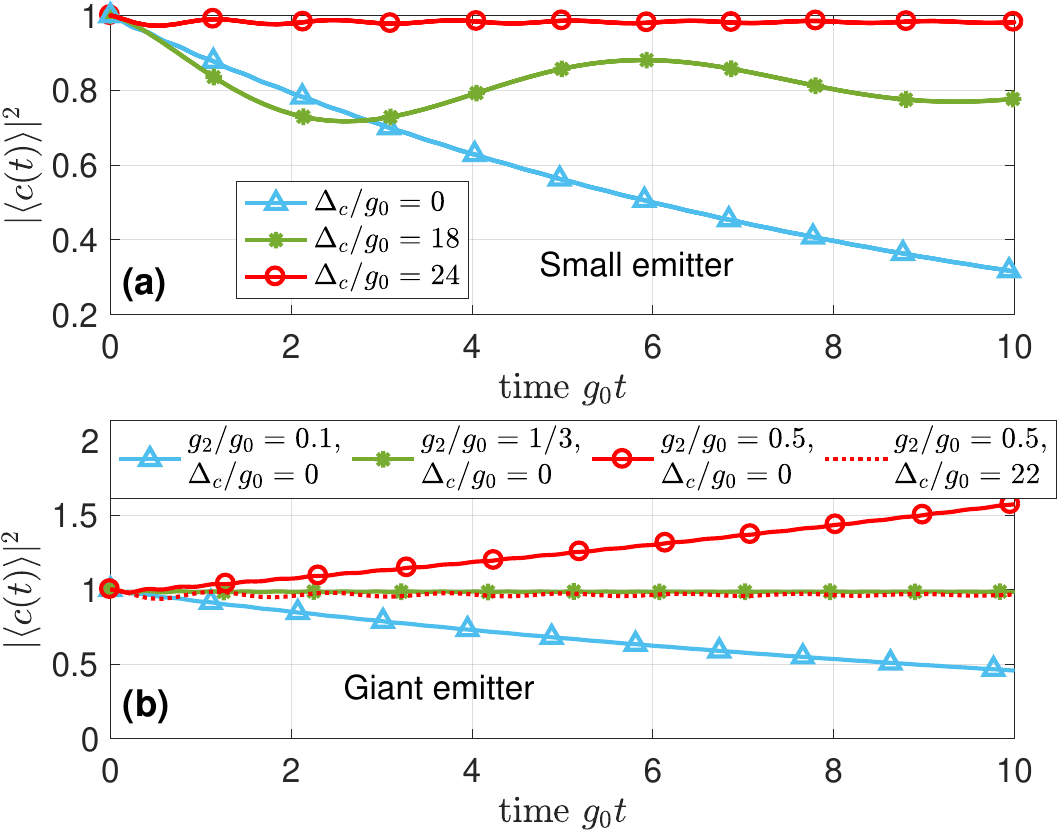}
\caption{Time evolutions of occupation $|\langle c(t)\rangle|^{2}$ of the quantum emitter $c$ in (a) the small-atom case with $g_{N}=0$ and (b) the giant-atom case with $g_{N}\neq0$ and $N=2$. Other parameters, except for those given in the legends, are $J/g_{0}=10$, $\gamma/g_{0}=5$, and $M_{\text{tot}}=800$.}\label{SingleE}
\end{figure}

We study the dynamics of the mean amplitudes $\langle c\rangle$ of the emitter and $\langle a_{j}\rangle$ of the lattice sites by solving the Heisenberg equations of motion 
\begin{eqnarray}
\langle\dot{c}\rangle&=&-i\Delta_{c}\langle c\rangle-i\mleft(g_{0}\langle a_{0}\rangle+g_{N}\langle a_{N}\rangle\mright), \label{EOMc}\\
\langle\dot{a}_{j}\rangle&=&-\kappa\langle a_{j}\rangle-i\mleft(J_{R}\langle a_{j-1}\rangle+J_{L}\langle a_{j+1}\rangle\mright)\nonumber\\
&&-i\mleft(g_{0}\delta_{j,0}+g_{N}\delta_{j,N}\mright)\langle c\rangle, \label{EOMa}
\end{eqnarray}
where $\kappa$ accounts for the (practically unavoidable) on-site energy loss of the HN model. At the quantum level, environmental fluctuations can lead to the breakdown of the non-Hermitian skin effect~\cite{JonasPRR2024}. In the above equations, we have omitted the intrinsic dissipation of the emitter since it can typically be made very small (compared to other characteristic scales) in many platforms. Nevertheless, as will be discussed below, the intrinsic dissipation of a small emitter inevitably leads to a trivial damping of the dynamics, while for a giant emitter it can be offset under specific conditions. 

%%%%%%%%%%%%%%%%%%%%%%%%%%%%%%%%%%%%%%%%%%%%%%%%%%%%%%

\subsection{Dynamics of a single emitter}
\label{secS}

We first study the dynamics of a single emitter, i.e., the emitter $c$ introduced above, in both the small- and giant-emitter cases. For a Hermitian lattice, it is known that a small emitter shows complete (fractional) decay, i.e., the emitter will eventually lose all (a fraction of) the excitation, if its frequency is within the energy band (band gap) of the lattice. This behavior is slightly modified, however, for the non-Hermitian HN model considered here. In this case, the \emph{effective} band edge is determined by the hopping rate $\sqrt{J_{R}J_{L}}$ of the pseudo-Hermitian model in Eq.~(\ref{mapH}) rather than the maximum real part $|J_{R}|$ of the spectrum~\cite{LonghiQD}. As shown in Fig.~\figpanel{SingleE}{a}, even though the lattice enables directional field amplification, the small emitter with $g_{N}=0$ shows complete (fractional) decay when $|\Delta_{c}|<2\sqrt{J_{R}J_{L}}$ ($|\Delta_{c}|>2\sqrt{J_{R}J_{L}}$), and the dynamics become almost dissipationless when $|\Delta_{c}|\gg2\sqrt{J_{R}J_{L}}$. The small emitter can never exhibit energy growth as long as $J_{R}$ and $J_{L}$ have the same sign. 

In the giant-emitter case, the dynamics are sensitive to the ratio $g_{N}/g_{0}$ of the two coupling strengths. Different from the Hermitian case, where a giant emitter can be decoherence-free (i.e., immune to radiating into the lattice) if, e.g., $\Delta_{c}=0$ and $N=2$~\cite{AFKstructured}, in the non-Hermitian case decoherence-free dynamics also require carefully matched coupling strengths~\cite{DLHN}. Otherwise, the giant emitter exhibits either complete decay or secular energy growth as shown in Fig.~\figpanel{SingleE}{b}. The matching condition for the coupling can be analytically identified by calculating the self-energy of the giant emitter (see Appendix~\ref{appSE} for more details), i.e., 
\begin{equation}
\begin{split}
\Sigma_{c}(z)&=\mp \frac{1}{\sqrt{z^2-4J_{R}J_{L}}}\\
&\quad\,\times\mleft[G_{0}^2+G_{N}^2+G_{0}G_{N}y_{\pm}^{N}\mleft(\beta^{N}+\beta^{-N}\mright)\mright],
\end{split}
\label{selfc}
\end{equation}
where $y_{\pm}=(z\pm\sqrt{z^2-4J_{R}J_{L}})/(2\sqrt{J_{R}J_{L}})$ and $G_{j}=g_{j}/\sqrt{2\pi}$. The upper or lower sign is selected depending on whether $y_{+}$ or $y_{-}$ is located within the unit circle in the complex plane~\cite{AFKstructured}. 

The dynamics in Fig.~\figpanel{SingleE}{b} can be well understood from $\Sigma_{c}(\Delta_{c}+i0^{+})$, whose real and imaginary parts capture the frequency shift and decay rate of the emitter induced by the lattice~\cite{AGT2017pra,AGT2017prl}, respectively. Clearly, one has $\Sigma_{c}(0+i0^{+})=0$ if, e.g.,
\begin{equation}
\frac{g_{N}}{g_{0}}=\beta^{\pm N}, \quad\, \text{mod}(N,4)=2,
\label{matchcondition}
\end{equation}
which is exactly the condition corresponding to the decoherence-free dynamics (green line with asterisks) in Fig.~\figpanel{SingleE}{b}. The first half of Eq.~(\ref{matchcondition}) ensures the balance between the emitted and absorbed energies at the two coupling points, a prerequisite for achieving perfect destructive interference, while the second half, in the ideal limit, enables the destructive interference between the local decay channels [terms proportional to $G_{0}^{2}$ and $G_{N}^{2}$ in Eq.~(\ref{selfc})] and the nonlocal (cooperative) decay channels [terms proportional to $G_{0}G_{N}$ in Eq.~(\ref{selfc})], thereby leading to effective light-matter decoupling. Note that in the non-Hermitian case, $\Sigma_{c}(0+i0^{+})$ can have a \emph{negative} imaginary part, which accounts for the energy growth of the emitter (red lines with circles). Moreover, the giant emitter shows similar dissipationless dynamics when its frequency completely falls within the band gap, similar to the small-emitter case. 

\begin{figure}[ptb]
\centering
\includegraphics[width=\linewidth]{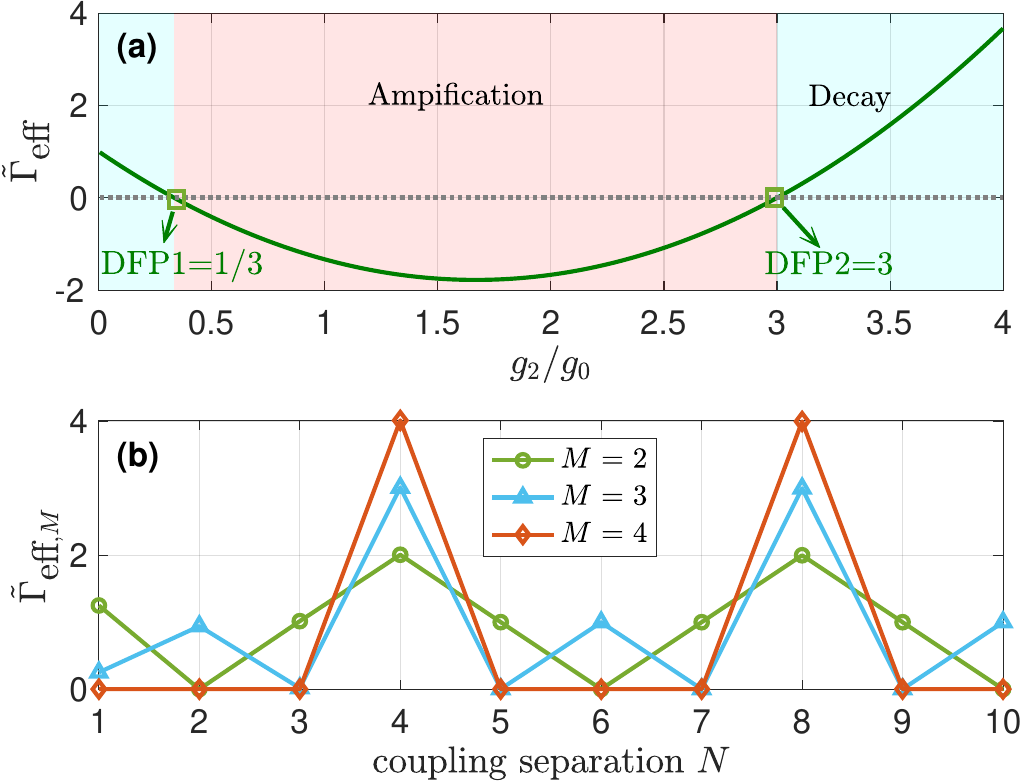}
\caption{(a) Dimensionless effective decay rate $\tilde{\Gamma}_{\text{eff}}$ of a resonant giant emitter with two coupling points as a function of the coupling ratio $g_{2}/g_{0}$, with the coupling separation $N=2$. (b) Dimensionless effective decay rate $\tilde{\Gamma}_{\text{eff},M}$ of a resonant giant emitter with $M$ equally spaced coupling points as a function of the coupling-point separation $N$. In panel (a), the gray dotted line indicates the position of zero decay rate, while the cyan and red areas indicate the regions of complete decay and secular energy growth, respectively. In panel (b), we consider a coupling matching condition $g_{jN}/g_{j'N}=\beta^{-|j-j'|N}$ ($0\leq j'<j\leq M-1$) for all coupling points with $\beta=2$.}
\label{Edecay}
\end{figure}

From the self-energy in Eq.~(\ref{selfc}) one can find another interesting result: there are at most \emph{two} decoherence-free points (DFPs) where the giant emitter shows dissipationless dynamics. In Fig.~\ref{Edecay}, we plot the dimensionless effective decay rate for the above resonant giant-emitter case, which is defined as
\begin{equation}
\tilde{\Gamma}_{\text{eff}}=\frac{2\text{Im}[\Sigma_{c}(0+i0^{+})]\sqrt{J_{R}J_{L}}}{G_{0}^{2}}.
\label{effectivedecay}
\end{equation}
It is clear that emitter $c$ shows complete decay (with a positive $\tilde{\Gamma}_{\text{eff}}$) if the coupling ratio $g_{2}/g_{0}$ is smaller (larger) than the first (second) DFP, while otherwise it shows a secular energy growth (with a negative $\tilde{\Gamma}_{\text{eff}}$). At the two DFPs, the effective decay rate vanishes such that the giant emitter is immune to energy relaxation or amplification. We would like to point out that the amplification regime is exclusive to giant emitters and plays a crucial role in preserving the decoherence-free dynamics from extra dissipations of the system, as will be discussed in Sec.~\ref{secEnhance}.

\begin{figure}[ptb]
\centering
\includegraphics[width=\linewidth]{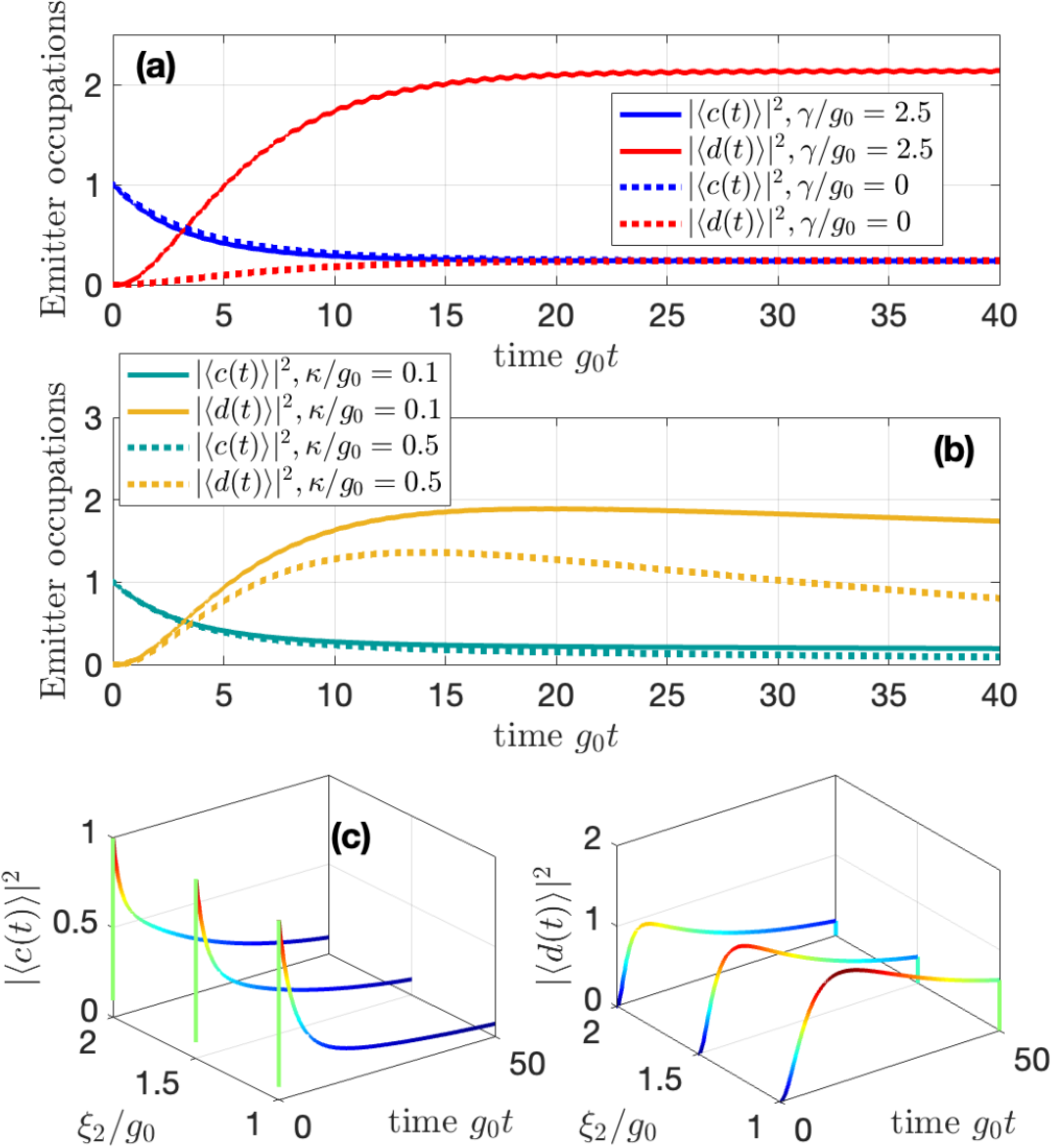}
\caption{Dynamics in the two-small-atoms case. Time evolutions of occupations $|\langle c(t)\rangle|^{2}$ and $|\langle d(t)\rangle|^{2}$ with (a) different values of coupling imbalance $\gamma$, (b) different values of on-site loss rate $\kappa$, and (c) different values of coupling ratio $\xi_{3}/g_{0}$. We assume $\kappa=0$ and $\xi_{2}/g_{0}=1$ in (a), $\gamma/g_{0}=2.5$ and $\xi_{2}/g_{0}=1$ in (b), and $\kappa/g_{0}=0.5$ and $\gamma/g_{0}=2.5$ in (c). Other parameters are $J/g_{0}=5$, $\Delta_{c}=\Delta_{d}=0$, and $M_{\text{tot}}=800$.}\label{Dark}
\end{figure}

Before addressing multi-emitter cases, we briefly discuss how the number of coupling points influences the results presented above (so far we have considered only two coupling points for the giant emitter). Without loss of generality, we consider a simplified case where all coupling points are equally spaced (e.g., with a separation of $N$ lattice sites between adjacent coupling points) and maintain the same coupling matching condition as above, e.g., $g_{jN}/g_{j'N}=\beta^{-|j-j'|N}$ for $0\leq j'<j \leq M-1$. In this case, the dimensionless effective decay rate of a resonant giant emitter with $M$ coupling points can be expressed as
\begin{eqnarray}
\tilde{\Gamma}_{\text{eff},M}&=&\text{Re}\Bigg[\sum_{j,j'=0}^{M-1}(\pm i)^{|j-j'|N}\beta^{-(j+j')N}\nonumber\\
&&\times\mleft(\beta^{|j-j'|N}+\beta^{-|j-j'|N}\mright)\Bigg].
\label{effectivedecay2}
\end{eqnarray}
Note that the coupling matching condition considered above, although convenient for analytical simplification, is not strictly required when considering more than two coupling points. This is somewhat similar to the Hermitian case, where a giant atom with more than two coupling points can still be effectively decoupled from the lattice even if the coupling strengths are not equal.

Figure~\figpanel{Edecay}{b} shows the effective decay rate $\tilde{\Gamma}_{\text{eff},M}$ as a function of the coupling separation $N$ for different numbers of coupling points. One can find that the giant emitter always exhibits a periodic effective decay as $N$ varies, similar to the Hermitian case. Notably, increasing the number of coupling points can introduce additional decoherence-free points and trivially enhance the maximum of the effective decay rate (due to more coupling points). Nevertheless, the interplay between the self-interference effect of the giant emitter and the non-Hermitian skin effect of the lattice remains qualitatively unchanged. Therefore, hereafter we will focus on giant emitters with two coupling points for simplicity.

%%%%%%%%%%%%%%%%%%%%%%%%%%%%%%%%%%%%%%%%%%%%%%%%%%%%%%

\subsection{Dynamics of two small emitters}
\label{secTwoS}

\begin{figure*}[ptb]
\centering
\includegraphics[width=0.95\linewidth]{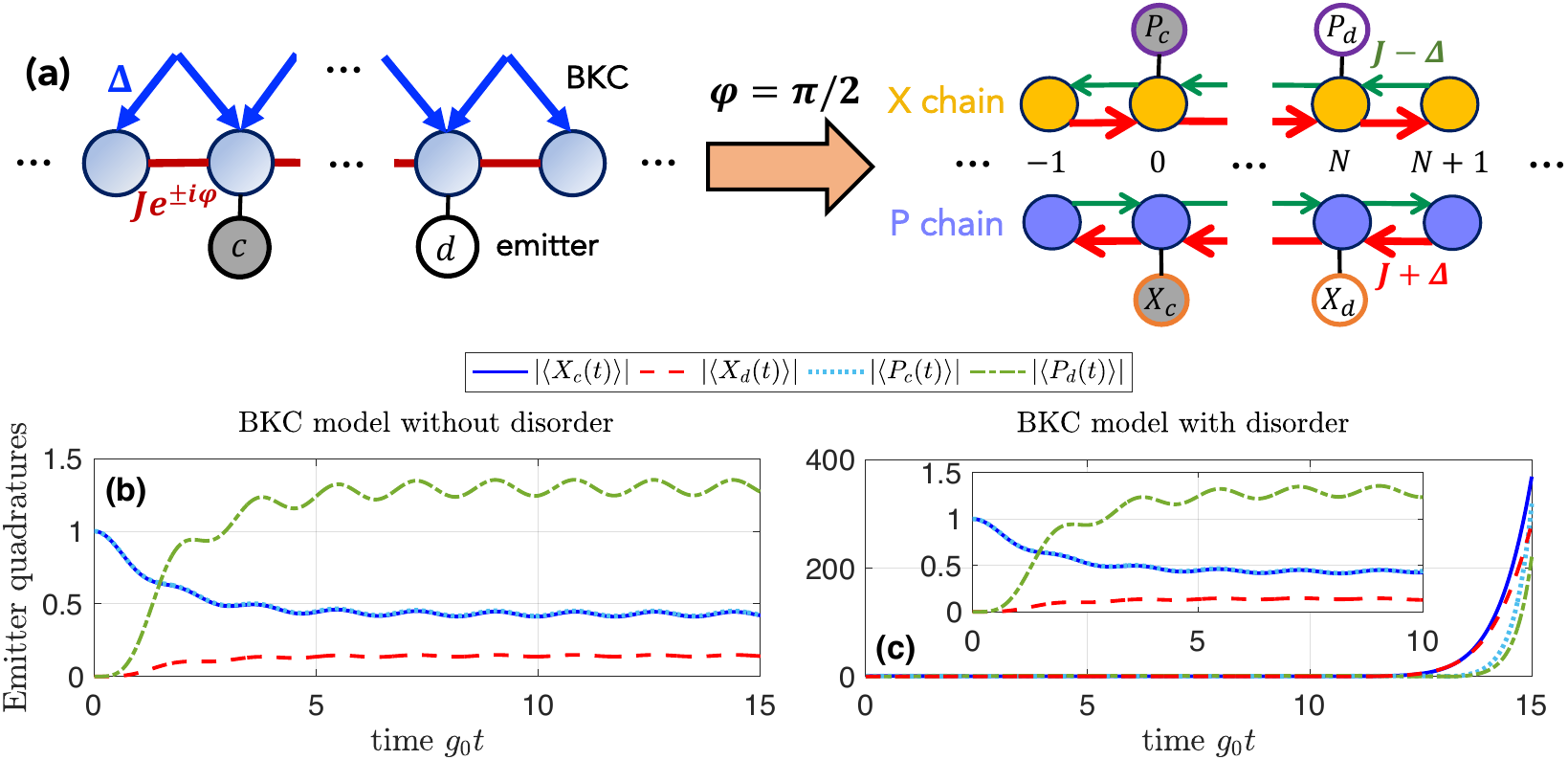}
\caption{(a) Schematic diagram of two (small) quantum emitters coupled to a bosonic Kitaev chain (BKC). The BKC features a nearest-neighbor hopping term with amplitude (phase) $J$ ($\varphi$) and a pairing potential with strength $\Delta$. When $\varphi=\pi/2$, the position (momentum) quadratures $X_{c,d}$ ($P_{c,d}$) of the two emitters are coupled to a Hatano--Nelson chain composed of the momentum (position) quadratures of the BKC site modes. The two effective Hatano--Nelson chains (i.e., the $X$ and $P$ chains) are opposite. (b, c) Time evolutions of emitter quadratures $|X_{c,d}(t)|$ and $|P_{c,d}(t)|$ (b) without (i.e., $W_{p}/g_{0}=0$) and (c) with (i.e., $W_{p}/g_{0}=10^{-8}$) on-site potential disorder. Other parameters are $J/g_{0}=2$, $\gamma/g_{0}=1$, $\Delta_{c}=\Delta_{d}=0$, $\xi_{2}/g_{0}=1$, and $M_{\text{tot}}=800$.}\label{Squeeze}
\end{figure*}

Now we consider an additional quantum emitter $d$, coupled to lattice sites $N'$ and $N''$ with coupling strengths $\xi_{N'}$ and $\xi_{N''}$, respectively, and study the unconventional inter-emitter interactions mediated by the HN model. Similar to emitter $c$, we include in the total Hamitonian the free-energy term $\Delta_{d}d^{\dag}d$ of emitter $d$ and the corresponding interaction terms $(\xi_{N'}a_{N'}+\xi_{N''}a_{N''})d^{\dag}+\text{H.c.}$. 

We first focus on the small-emitter case where both $c$ and $d$ are coupled locally to the HN model (i.e., $g_{N}=\xi_{N''}=0$). In this case, a very interesting result is that under specific conditions, the two emitters will evolve into a long-lived superposition state, with their final occupations determined by the degree of the non-Hermiticity. As shown in Fig.~\figpanel{Dark}{a}, while in the Hermitian case the two emitters exhibit identical final occupations if $N'=2$ and $g_{0}=\xi_{2}$, in the non-Hermitian case the right emitter $d$ exhibits a larger final occupation than the left emitter $c$ under the same condition. This can be clearly understood from the fact that the lattice field is amplified towards the right. Note that in this case the amplification direction of the emitter's excitation is \emph{identical} to that of the field. This is very different from the situation where giant emitters exhibit unconventional (i.e., nonreciprocal) DFI in the non-Hermitian case, as will be discussed in Sec.~\ref{secTwoG}. 

Up to now we have focused on the situation where the HN model is completely lossless. In practice, however, the lattice sites will also be subjected to unavoidable energy losses. This dissipation effect is described by the on-site decay term $-i\kappa \langle a_{j}\rangle$ in Eq.~(\ref{EOMa}). In Fig.~\figpanel{Dark}{b}, we show that the long-lived superposition state can be immediately broken in the presence of the loss, with the damping rate of the emitter occupations depending on the loss rate $\kappa$. This damping cannot be avoided by tuning the coupling ratio $\xi_{2}/g_{0}$, as shown in Fig.~\figpanel{Dark}{c}. Instead, changing $\xi_{2}/g_{0}$ only affects the (maximum) occupation of emitter $d$. This is very different from the decoherence-free behavior of giant emitters, which, as will be discussed in Sec.~\ref{secEnhance}, persists even in the presence of such extra dissipation. This difference can be understood from the fact that small emitters can never show energy amplification (as long as $J_{R}$ and $J_{L}$ have the same sign), which is able to balance the extra dissipation of the system.

The above result has interesting extensions; for instance, if we introduce a more advanced lattice model with a \emph{phase-sensitive} non-Hermitian skin effect. More specifically, we consider a bosonic analogue of the celebrated fermionic Kitaev model~\cite{FermKitaev}, which is described by the Hamiltonian
\begin{equation}
%\begin{split}
H_{\text{BKC}}=\sum_{j}\mleft(Je^{i\varphi}a_{j+1}^{\dag}a_{j}+i\Lambda a_{j+1}^{\dag}a_{j}^{\dag}+\text{H.c.}\mright),
\label{BKCH}
%\end{split}
\end{equation}
where $J$ and $\varphi$ are the amplitude and phase, respectively, of the nearest-neighbor hopping coefficient; $\Lambda$ is the strength of the nearest-neighbor ``two-mode-squeezing'' interaction, mimicking the p-wave pairing potential in its fermionic version. 

This model is known as the \emph{bosonic Kitaev chain} (BKC)~\cite{Aash2018prx, Aash2020nc, Chris2023arxiv, Clara2024nature}. It enables phase-sensitive directional amplifications if $\varphi=\pi/2$ (modulo $\pi$). As shown in Fig.~\figpanel{Squeeze}{a}, this feature can be better understood if we define position $x_{j}=(a_{j}^{\dag}+a_{j})/\sqrt{2}$ and momentum $p_{j}=i(a_{j}^{\dag}-a_{j})/\sqrt{2}$ quadratures for the lattice sites and write down their equations of motion in the case of $\varphi=\pi/2$:
\begin{eqnarray}
\dot{x}_{j}&=&\mleft(J+\Delta\mright)x_{j-1}-\mleft(J-\Delta\mright)x_{j+1}, \label{xEOM}\\
\dot{p}_{j}&=&\mleft(J-\Delta\mright)p_{j-1}-\mleft(J+\Delta\mright)p_{j+1}. \label{pEOM}
\end{eqnarray}
Clearly, the position and momentum quadratures form two \emph{independent and opposite} HN models, such that a wave packet can be directionally amplified towards different directions depending on its initial phase~\cite{Aash2018prx}. If we further introduce two (small) quantum emitters and define similar orthogonal quadratures $X_{o}=(o^{\dag}+o)/\sqrt{2}$ and $P_{o}=i(o^{\dag}-o)/\sqrt{2}$ ($o=c,\,d$) for them, as shown in Fig.~\figpanel{Squeeze}{a}, we can easily find that the momentum (position) quadratures of the emitters are coupled to the position (momentum) quadrature HN chain of the lattice sites (see Appendix~\ref{appBKC} for more details).  

As shown in Fig.~\figpanel{Squeeze}{b}, if two resonant small emitters (i.e., $\Delta_{c}=\Delta_{d}=0$) are coupled to the BKC model with coupling separation $N=2$, one can expect state transfer between these two emitters. The initially unoccupied emitter (i.e., emitter $d$ here) will eventually exhibit different occupations in its position and momentum quadratures since the two effective HN chains are opposite, such that the excitation is amplified (attenuated) when transferring from $P_{c}$ to $P_{d}$ (from $X_{c}$ to $X_{d}$). That is to say, while emitter $c$ always maintains the quadrature balance (i.e., $X_{c}\equiv P_{c}$), in emitter $d$ most of the excitation is ``squeezed'' into the momentum quadrature, with the quadrature imbalance determined by both the emitter-lattice coupling strengths and the non-Hermiticity.  

We point out, however, that the dynamics of the emitters are \emph{extremely fragile} to on-site potential disorders of the BKC. This is due to the fact that the two quadrature HN chains of the BKC model are no longer decoupled in the presence of on-site potential disorders~\cite{Aash2018prx}:
\begin{eqnarray}
\dot{x}_{j}&=&\mleft(J+\Delta\mright)x_{j-1}-\mleft(J-\Delta\mright)x_{j+1}+\delta_{j}p_{j}, \label{xEOMdis}\\
\dot{p}_{j}&=&\mleft(J-\Delta\mright)p_{j-1}-\mleft(J+\Delta\mright)p_{j+1}-\delta_{j}x_{j}, \label{pEOMdis}
\end{eqnarray}
where $\delta_{j}\in[-W_{p},\,W_{p}]$ is the on-site potential of the $j$th lattice site with $W_p$ the maximum disorder strength. In this case, the field can travel back and forth between the two opposite HN chains, thus leading to indefinite amplification and instability. As shown in Fig.~\figpanel{Squeeze}{c}, the emitters always exhibit secular energy growth at long times, even if only an extremely small disorder of $W_{p}/g_{0}=10^{-8}$ is introduced. This is quite different from the end-to-end \emph{photon scattering} of the BKC, which can remain robust against weak enough on-site disorders if the chain is sufficiently short~\cite{Aash2018prx}. Nevertheless, the inset in Fig.~\figpanel{Squeeze}{c} shows that for weak disorders, one can still expect an initial transient stage where emitter $d$ exhibits different occupations in its position and momentum quadratures, before significant energy growth appears.

\begin{figure}[ptb]
\centering
\includegraphics[width=\linewidth]{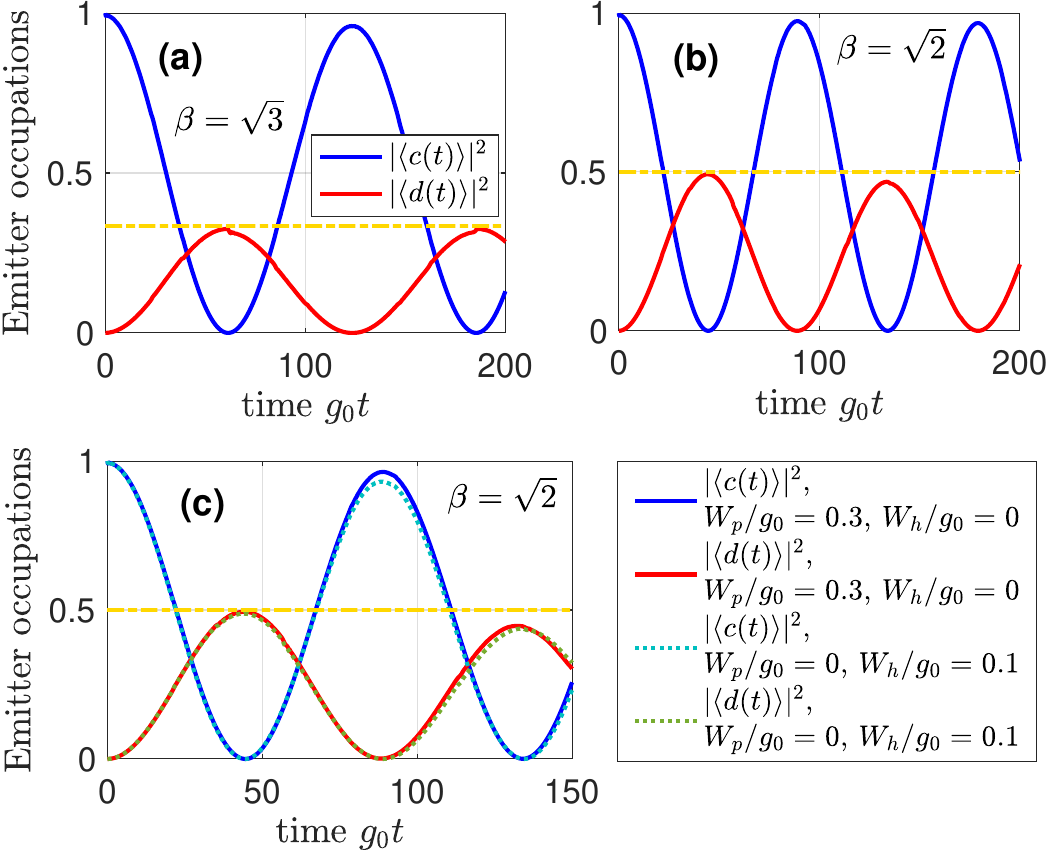}
\caption{Dynamics in the two-giant-atoms case. (a) Time evolutions of occupations $|\langle c(t)\rangle|^{2}$ and $|\langle d(t)\rangle|^{2}$ for an ideal HN model with (a) $\beta=\sqrt{3}$ (i.e., $\gamma/g_{0}=7.5$) and (b) $\beta=\sqrt{2}$ (i.e., $\gamma/g_{0}=5$). Panels (a) and (b) share the same legend. (c) Time evolutions of occupations $|\langle c(t)\rangle|^{2}$ and $|\langle d(t)\rangle|^{2}$ for an HN model with $\beta=\sqrt{2}$ (i.e., $\gamma/g_{0}=5$) and different types of disorder. In all panels, the yellow dotted-dashed lines indicate the values of $\beta^{2}$. Other parameters, except for those given in the legends, are $J/g_{0}=15$, $\Delta_{c}=\Delta_{d}=0$, $\xi_{1}/g_{0}=1$, $g_{2}/g_{0}=\xi_{3}/g_{0}=1/\beta^{2}$, and $M_{\text{tot}}=800$.}\label{DFI}
\end{figure}

The above result is a simple example that reveals the dynamical instability arising from the secular amplification of the fields propagating in the lattice. In fact, similar results can occur as long as the amplified emission fields have an opportunity to return to the emitters, such as in the HN model under periodic boundary conditions~\cite{DLHN}. This feature is closely related to the sensitivity of the non-Hermitian skin effect to boundary conditions, which has no Hermitian counterparts. Nevertheless, unlike the HN model under periodic conditions, the field amplification in a disordered BKC is completely indefinite since the disorders, which determine the complicated inter-chain couplings, are random.

%%%%%%%%%%%%%%%%%%%%%%%%%%%%%%%%%%%%%%%%%%%%%%%%%%%%%%

\subsection{Dynamics of two giant emitters}
\label{secTwoG}

Now we turn to study the giant-emitter case, where each of the emitters $c$ and $d$ is coupled to two lattice sites. With a Hermitian 1D tight-binding chain, it has been shown that two well-arranged ``braided'' giant emitters can exhibit decoherence-free interactions (DFIs) if the non-Markovian retardation effect is negligible~\cite{AFKstructured}. In the non-Hermitian case, as shown in Fig.~\ref{DFI}, the DFI becomes \emph{nonreciprocal} and the nonreciprocity is opposite to that of the field~\cite{DLHN}. Such nonreciprocal DFIs are realized when the decoherence-free condition Eq.~(\ref{matchcondition}) is fulfilled for each of the emitters and their coupling points are braided (i.e., $N''>N'>N>0$). In view of this, here we take $g_{0}=\xi_{1}$ and $g_{2}=\xi_{3}=g_{0}/\beta^{2}$ in Fig.~\ref{DFI} as an example. It is clear from Figs.~\figpanel{DFI}{a} and \figpanel{DFI}{b} that the nonreciprocity is quantitatively determined by $\beta$: the excitation is attenuated by $\beta^{-2}$ when it is transferred from $c$ to $d$. Such a nonreciprocal DFI can be used to realize non-Hermitian cooling~\cite{NHcooling}, i.e., the steady-state thermal occupation of one emitter is smaller than that of the other, and the cooling effect can be exponentially enhanced by integrating more (giant) emitters in this form. 

The nonreciprocal DFIs can be explicitly understood from the interaction parts (i.e., the off-diagonal elements) of the $2\times2$ level-shift operator of the emitter pair, which, under the above parametric conditions, can be obtained as (see Appendix~\ref{appDE} for more details)
\begin{eqnarray}
\Sigma_{cd}(0+i0^{+})&\simeq&\frac{-G_{0}^{2}}{J_{R}}, \label{selfcd}\\
\Sigma_{dc}(0+i0^{+})&\simeq&\frac{-G_{0}^{2}}{J_{R}\beta^{2}}. \label{selfdc}
\end{eqnarray}
Similar to the self-energy in Eq.~(\ref{selfc}), the real and imaginary parts of $\Sigma_{cd}$ ($\Sigma_{dc}$) describe the \emph{coherent} and \emph{dissipative} couplings from $d$ to $c$ (from $c$ to $d$), respectively. The effective interaction between the two giant emitters is purely coherent, and the coupling nonreciprocity, i.e., $|\Sigma_{cd}(0+i0^{+})/\Sigma_{dc}(0+i0^{+})|=\beta^{2}$, is exactly determined by the non-Hermiticity of the HN model and the separations of the coupling points. 

Besides the intrinsic dissipations discussed above, the HN model also suffers from unavoidable disorders of the hopping rates and on-site potentials in practice. At first glance, the nonreciprocal DFIs are implemented by judiciously matching the emitter-lattice coupling coefficients with the global non-Hermiticity of the lattice, and should thus be sensitive to these disorders. However, we reveal in Fig.~\figpanel{DFI}{c} that the DFIs are actually robust against moderate disorders, which only introduce minor damping and shifts to the evolution curves. This robustness is reminiscent of how the HN model under open boundary conditions remains in the \emph{skin phase} for weak disorders~\cite{LonghiDisorder2021}. Here the on-site potential disorder is defined in the same way as in Fig.~\figpanel{Squeeze}{c}, while the hopping disorder is introduced by assuming position-dependent hopping differences $\nu_{j}=J_{R,j}-J_{L,j}\in[-W_{h},\,W_{h}]$ with $J_{R,j}$ ($J_{L,j}$) the rightward (leftward) hopping rate connecting the $j$th and $j+1$th lattice sites. Compared with the hopping disorder, the dynamics are more robust against the on-site potential disorder (note that we have assumed a larger strength $W_{p}$ for the on-site potential disorder). This is consistent with the fact that the nonreciprocal DFI is associated with the asymmetric hopping rates rather than the on-site terms. 

%%%%%%%%%%%%%%%%%%%%%%%%%%%%%%%%%%%%%%%%%%%%%%%%%%%%%%

\subsection{Giant emitters coupled to a bosonic Kitaev chain}
\label{secGABKC}

\begin{figure}[ptb]
\centering
\includegraphics[width=\linewidth]{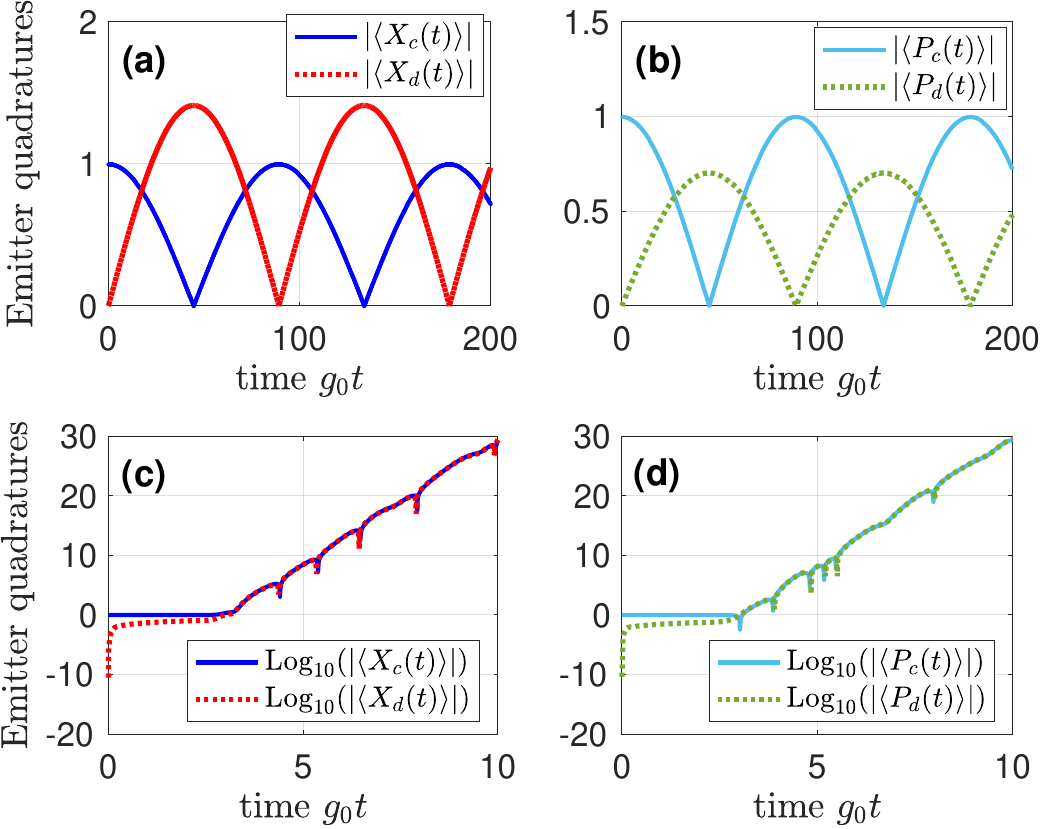}
\caption{Time evolutions of the position $|X_{c,d}(t)|$ and momentum $|P_{c,d}(t)|$ emitter quadratures of two braided giant emitters coupled to a BKC (a, b) without and (c, d) with on-site potential disorder. We assume $W_{p}/g_{0}=0$ in panels (a) and (b) and $W_{p}/g_{0}=10^{-5}$ in panels (c) and (d). Other parameters are $J/g_{0}=15$, $\gamma/g_{0}=5$, $\Delta_{c}=\Delta_{d}=0$, $\xi_{1}/g_{0}=1$, $g_{2}/g_{0}=\xi_{3}/g_{0}=1/\beta^{2}$, and $M_{\text{tot}}=800$.}\label{GABKC}
\end{figure}

Similar to the small-emitter case, an ideal BKC is also supposed to be a promising platform for enabling phase-sensitive interactions between giant emitters. More specifically, we consider two identical giant emitters $c$ and $d$ (as in Sec.~\ref{secTwoG}) coupled to a BKC with the interaction Hamiltonian
\begin{equation}
H_{\text{int}}=(g_{0}a_{0}+g_{2}a_{2})c^{\dag}+(\xi_{1}a_{1}+\xi_{3}a_{3})d^{\dag}+\text{H.c.},
\end{equation}
where $a_{j}$ denotes the annihilation operator of the $j$th lattice site of the BKC in this case. The emitter-lattice coupling strengths are matched as in Fig.~\ref{DFI}. Once again, according to Eqs.~(\ref{SMnewx})--(\ref{SMEOMpc}) in Appendix~\ref{appBKC}, the position and momentum quadratures of each giant emitter are coupled to the P chain and X chain, respectively, with identical coupling separation and coupling strengths. Consequently, the two braided giant emitters support two pairs of quadratures, which are respectively coupled to two opposite HN chains with the same braided coupling structure. 

As shown in Figs.~\figpanel{GABKC}{a} and \figpanel{GABKC}{b}, the position $|X_{c,d}(t)|$ and momentum $|P_{c,d}(t)|$ quadratures of the giant emitters exhibit opposite nonreciprocal DFIs, which we attribute to the opposite chirality of the two effective HN chains. This implies that the BKC can, in the ideal limit, support \emph{phase-sensitive} DFIs with the nonreciprocity depending on the initial phase of the excitation. Different from the result in Fig.~\ref{DFI}, here the amplification (attenuation) factor is $\beta$ ($\beta^{-1}$) since we plot the amplitudes of the emitter quadratures rather than their intensities.   

As in the small-emitter case, such an intriguing phenomenon is, however, disrupted by the unavoidable on-site potential disorders of the BKC. As shown in Figs.~\figpanel{GABKC}{c} and \figpanel{GABKC}{d}, the dynamics in the ideal case can be completely destroyed by extremely weak disorders (similar results can be obtained even with much weaker disorders, e.g., $W_{p}/g_{0}=10^{-20}$), showing instead secular growth in the long-time limit. Different from the small-emitter case, where one can still expect transient phase-sensitive dynamics before the energy growth dominates, the timescale of the nonreciprocal DFI is typically much larger than the onset time of the energy growth. Therefore, the phase-sensitive DFIs shown in Figs.~\figpanel{GABKC}{a} and \figpanel{GABKC}{b} can hardly be observed in practice. 

\begin{figure}[ptb]
\centering
\includegraphics[width=\linewidth]{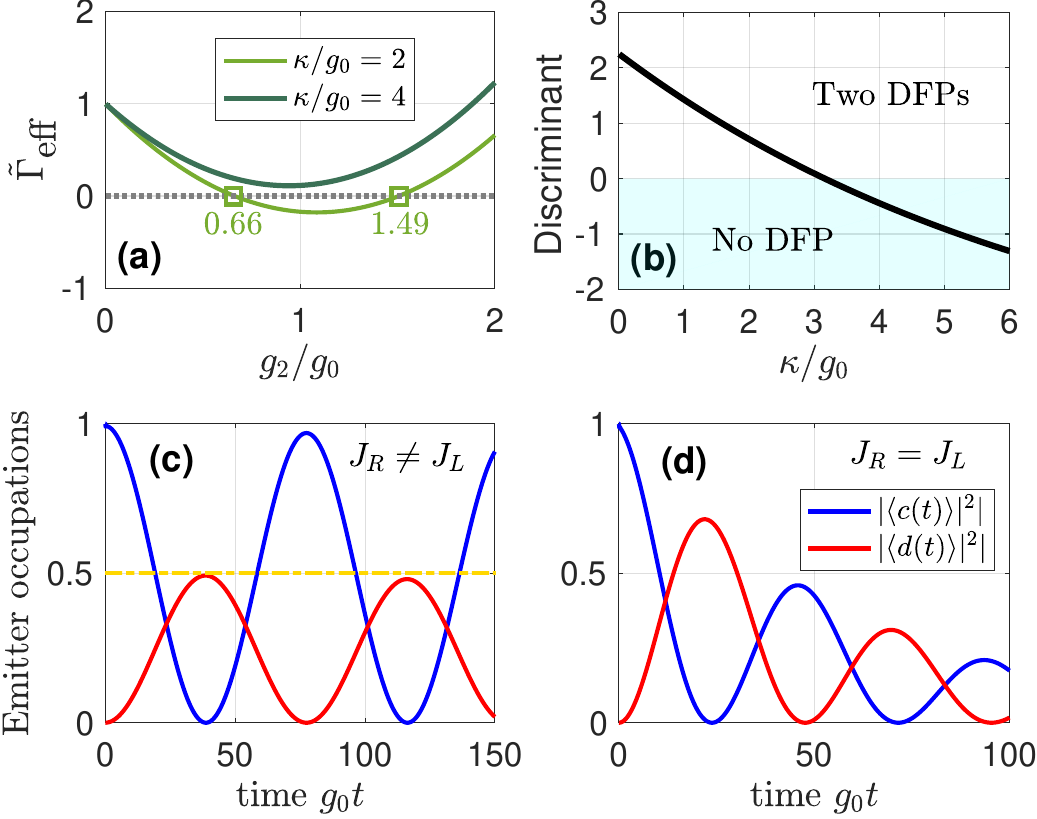}
\caption{(a) Dimensionless effective decay rate $\tilde{\Gamma}_{\text{eff}}$ of a resonant giant emitter as a function of the coupling ratio $g_{2}/g_{0}$ ($N=2$) in the presence of on-site energy loss and for $\gamma/g_{0}=5$. The gray dotted line indicates the position of zero. (b) The discriminant in Eq.~(\ref{discriminant}) as a function of on-site loss rate $\kappa$ for $N=2$ and $\gamma/g_{0}=5$. The cyan area indicates the region where the discriminant is negative (with no DFP). (c, d) Time evolutions of occupations $|\langle c(t)\rangle|^{2}$ and $|\langle d(t)\rangle|^{2}$ of the two giant emitters (c) with and (d) without the non-Hermitian skin effect and for $\kappa/g_{2}=2$. We assume $\gamma/g_{0}=5$ and $g_{2}/g_{0}=\xi_{3}/g_{0}=0.66$ in (c) and $\gamma/g_{0}=0$ and $g_{2}/g_{0}=\xi_{3}/g_{0}=1$ in (d). The yellow dot-dashed line in (c) indicates the value of $\beta^{2}$. Panels (c) and (d) share the same legend. Other parameters, except for those given in the legends, are $J/g_{0}=15$, $\Delta_{c}=\Delta_{d}=0$, $\xi_{1}/g_{0}=1$, and $M_{\text{tot}}=800$.}\label{Loss}
\end{figure}

%%%%%%%%%%%%%%%%%%%%%%%%%%%%%%%%%%%%%%%%%%%%%%%%%%%%%%

\subsection{Non-Hermiticity-enhanced decoherence-free dynamics}
\label{secEnhance}

Since the giant-emitter dynamics discussed in Sec.~\ref{secTwoG} are closely related to the non-Hermiticity of the bath, one may wonder how these results are influenced by intrinsic dissipations of the system, which appear as another non-Hermitain source. Clearly, the presence of the on-site energy loss can shift the spectrum of the HN model along the imaginary axis in the complex plane and thus modify its non-Hermiticity. Therefore the coupling matching condition $g_{0}/g_{N}=-\beta^{\pm N}$ introduced above will no longer apply. However, thanks to the amplification regime of a giant emitter (as shown in Fig.~\ref{Edecay}), the decoherence-free dynamics can be preserved even in the presence of moderate on-site energy losses. It can be analytically verified from Eq.~(\ref{selfc}) that the dimensionless effective decay rate $\tilde{\Gamma}_{\text{eff}}$ can still have zeros in this case. 

With finite on-site loss, the explicit form of the self-energy $\Sigma_{c}(z)$ (and thereby $\tilde{\Gamma}_{\text{eff}}$) should be modified by performing the transformation $z\rightarrow z+i\kappa$. In this case, we can identify the modified coupling matching condition by finding the value(s) of $g_{N}/g_{0}$ for which the effective decay rate of the giant emitter $c$ is zero, i.e.,
\begin{equation}
\text{Im}[\Sigma_{c}(\Delta_{c}+i\kappa+i0^{+})]=0.
\label{modifieddecay}
\end{equation}
However, as shown in Fig.~\figpanel{Loss}{a}, this condition cannot be fulfilled for a sufficiently strong on-site loss (e.g., $\kappa/g_{0}=4$). In that case, the effective decay rate is always positive so that the emitter exhibits complete decay in the long-time limit.
For the case in Fig.~\figpanel{Loss}{a}, where $\Delta_{c}=0$ and $N=2$, the condition in Eq.~(\ref{modifieddecay}) can be simplified to $\text{Re}[1+(g_{2}/g_{0})^{2}+y_{\pm}^{2}(g_{2}/g_{0})(\beta^{2}+\beta^{-2})]=0$. Clearly, the dimensionless effective decay rate $\tilde{\Gamma}_{\text{eff}}$ has two different zeros (no zero) if the discriminant
\begin{equation}
\left(\frac{i\kappa\pm i\sqrt{\kappa^{2}+4J_{R}J_{L}}}{2\sqrt{J_{R}J_{L}}}\right)^{4}\left(\beta^{2}+\beta^{-2}\right)-4
\label{discriminant}
\end{equation}
is positive (negative). We depict in Fig.~\figpanel{Loss}{b} the discriminant as a function of the on-site loss rate $\kappa$, which shows that the discriminant becomes negative for $\kappa/g_{0}\approx3.2$.

To verify the conclusion above, we examine in Fig.~\figpanel{Loss}{c} the dynamics of the two braided giant emitters with a finite $\kappa$. Here, the coupling ratio $g_{2}/g_{0}=0.66$ is (approximately) chosen as one of the two zeros of $\tilde{\Gamma}_{\text{eff}}$ as shown in Fig.~\figpanel{Loss}{a}. As expected, the nonreciprocal DFI is preserved under the new coupling matching condition. This is in sharp contrast to the conventional case without the non-Hermitian skin effect (i.e., $\gamma=0$ and $g_{0}=g_{2}=\xi_{1}=\xi_{3}$), as shown in Fig.~\figpanel{Loss}{d}, where the dynamics exhibit significant damping with the same value of $\kappa$. We thus conclude that the decoherence-free feature of a giant emitter can be enhanced by the non-Hermitian skin effect in the sense that the decoherence-free dynamics becomes more robust against moderate on-site energy loss of the structured bath. 

Finally, we point out that the HN model can also protect the decoherence-free dynamics from extra decay channels of the giant emitters (e.g., dissipation into the environment other than the HN mode) which typically poses a challenge to realize genuine decoherence-free (sub)-spaces. For instance, we have checked that a giant emitter with an extra decay rate $\Gamma_{\text{ext}}/g_{0}=0.02$ can still exhibit fractional decay for $\beta=\sqrt{3}$ and $g_{2}/g_{0}=1.26$. This condition is determined via balancing the extra decay of the giant emitter by the effective negative emitter decay induced by the HN model, i.e., 
\begin{equation}
\Gamma_{\text{ext}}+\text{Im}[\Sigma_{c}(\Delta_{c}+i0^{+})]=0. 
\label{balance}
\end{equation}
Equation~(\ref{balance}) clearly shows that the extra decay channels of the giant emitters have effects similar to those of the lattice. Specifically, the effective decay rate of the giant emitter increases with $\Gamma_{\text{ext}}$, and DFPs persist until $\Gamma_{\text{ext}}$ becomes sufficiently large. Once again, this ``gain-loss balance'' is not allowed in the absence of the non-Hermitian skin effect or in the small-emitter case.   

%%%%%%%%%%%%%%%%%%%%%%%%%%%%%%%%%%%%%%%%%%%%%%%%%%%%%%

\section{Conclusion and discussion}

In summary, we studied unconventional light-matter interactions between quantum emitters and the HN model, i.e., a minimal 1D structured bath that features the non-Hermitian skin effect, and revealed the interplay between the non-Hermiticity of the bath and various interference effects of the emitters. Our results showed that giant emitters can exhibit exclusive energy growth, which plays a crucial role in protecting the emitter dynamics from the unavoidable intrinsic dissipation of the system. Such protection is not possible in the absence of the non-Hermitian skin effect or with small emitters. We further showed that these dynamics are robust against several types of disorders, although some results seem to require well-matched parameters. However, unstable dynamics always appear when the field of the HN model can return to the emitters after being transferred away. This conclusion is verified with the BKC, which can be mapped to two opposite HN models under specific conditions but always exhibits dynamical instability in the presence of (infinitesimally small) on-site disorders. 

It is worth noting that the skin modes of the HN model, which inherit the nontrivial point-gap topology, show a \emph{topological} correspondence to two-dimensional integer quantum Hall states~\cite{Lee2019PRL}. However, these two-dimensional chiral modes still obey the conventional Bloch theorem, such that they are \emph{dynamically} distinct from the non-Hermitian skin modes which are responsible for the directional field amplification. Indeed, giant emitters show conventional self-interference effects (i.e., without requiring the coupling matching condition) when they are coupled to the chiral boundary states of a two-dimensional Harper-Hofstadter lattice~\cite{Topo2023Vega}.

The coupling matching condition considered in this paper is within reach of state-of-the-art experiments. In circuit quantum electrodynamics~\cite{Gu2017, Blais2021}, the coupling strength between quantum emitters, such as superconducting qubits and resonators, is typically determined by the design of the coupling elements, such as capacitors or inductors. The effective coupling strength can be tuned by adjusting parameters such as the size, spacing, or overlap of these elements~\cite{Coupler2021prapp}. Moreover, tunable and switchable coupling between two superconducting resonators can be mediated by an auxiliary coupler, such as a persistent current flux qubit~\cite{QubitCoupler2015} or a superconducting quantum interference device~\cite{SQUID2013prb}. Regarding the nonlocal coupling structures, there have been many proposals in which a single quantum emitter is able to interact with a lattice (e.g., a coupled-waveguide array) at multiple separate coupling points (see promising designs in, e.g., Refs.~\cite{LonghiGA,LKHpra2023}).

The results presented in this paper can inspire a series of intriguing investigations. For instance, introducing long-range (i.e., next-nearest-neighbor) asymmetric hoppings to the lattice can lead to bipolar (reciprocal) non-Hermitian skin effects~\cite{2021acousticNC}, which may mediate even more exotic interactions between quantum emitters. Another interesting direction is to study quantum emitters coupled to higher-dimensional extensions of the HN model~\cite{HigherOrderSkin, HigherOrderLee}, which exhibit skin modes localized at the edges and even corners. Such systems hold promise for engineering unconventional higher-dimensional interactions between quantum emitters. Moreover, many existing studies on quantum emitters coupled to topological lattices~\cite{BS2019Bello, Topo2018Barik, Topo2020GE, Topo2021Kim, Topo2021Vega, Topo2023Vega} can be straightforwardly extended to their non-Hermitian counterparts, where the interplay between the band topology and the non-Hermitian skin effect can play an important role in endowing the quantum emitters with additional unprecedented properties. It is also possible to modify this interplay by coupling different edge states of a non-Hermitian topological lattice through giant emitters~\cite{Competition2022Cheng}.

%%%%%%%%%%%%%%%%%%%%%%%%%%%%%%%%%%%%%%%%%%%%%%%%%%%%%%

\begin{acknowledgments}

L.D. thanks Christopher Wilson, Clara C. Wanjura, Francesco Ciccarello, Giuseppe La Rocca, Maurizio Artoni, Jin-Hui Wu, and Yao-Tong Chen for inspirations or discussions, and acknowledges financial support by the Knut och Alice Wallenberg stiftelse through project grant no. 2022.0090.

AFK acknowledges support from the Swedish Research Council (grant number 2019-03696), the Swedish Foundation for Strategic Research (grant numbers FFL21-0279 and FUS21-0063), the Horizon Europe programme HORIZON-CL4-2022-QUANTUM-01-SGA via the project 101113946 OpenSuperQPlus100, and from the Knut and Alice Wallenberg Foundation through the Wallenberg Centre for Quantum Technology (WACQT).

\end{acknowledgments}

%%%%%%%%%%%%%%%%%%%%%%%%%%%%%%%%%%%%%%%%%%%%%%%%%%%%%%

\appendix

\begin{widetext}

\section{Self-energy for a single giant emitter}
\label{appSE}

In this appendix, we provide a detailed derivation for the self-energy in Eq.~(7) in the main text. Before proceeding, we would like to show that the Hatano-Nelson (HN) model can be mapped to a psuedo-Hermitian lattice model, if it is in the \emph{convectively unstable regime} with $|J|>|\gamma|$~\cite{LonghiQD,DLHN}. Otherwise, if $|J|<|\gamma|$, the HN model enters the \emph{absolutely unstable regime}, where the quantum emitters always show secular energy growth regardless of the coupling strength.

In the convectively unstable regime, the Hamiltonian of the HN model can be rewritten as~\cite{Longhi2015prb}
\begin{equation}
H_{\text{HN}}'=\sum_{j}\mleft(\sqrt{J_{R}J_{L}}\beta a_{j+1}^{\dag}a_{j}+\sqrt{J_{R}J_{L}}\beta^{-1}a_{j}^{\dag}a_{j+1}\mright), 
\label{pesudoH}
\end{equation}
and its eigenvalues are given by
\begin{equation}
E_{q}=2\sqrt{J_{R}J_{L}}\cos{q},
\label{pesudoE}
\end{equation}
where $q=k-i\ln(\beta)=k-i\ln(\sqrt{J_{R}/J_{L}})$ is a \emph{complex} wave vector. This implies that the HN model is equivalent to a 1D pseudo-Hermitian lattice subject to an \emph{imaginary} gauge field. In view of this, by performing the transformation
\begin{equation}
a_{q}=\frac{1}{\sqrt{2\pi}}\sum_{n}a_{n}e^{-iqn},
\label{transform}
\end{equation}
the total Hamiltonian $H_{\text{tot}}$ of the system can be rewritten as
\begin{equation}
\tilde{H}_{\text{tot}}=H_{c}+\tilde{H}_{\text{HN}}+\tilde{H}_{\text{int}},
\label{kHtot}
\end{equation}
where
\begin{align}
H_{c}&=\Delta_{c}c^{\dag}c,
\label{HcSM}
\\
\tilde{H}_{\text{HN}}&=\sum_{q}\omega_{k}a_{q}^{\dag}a_{q}, \label{kHNH}
\\
\tilde{H}_{\text{int}}&=\sum_{q}\mleft[\mleft(G_{0}+G_{N}e^{-iNq}\mright)a_{q}^{\dag}c+\mleft(G_{0}+G_{N}e^{iNq}\mright)c^{\dag}a_{q}\mright] , \label{kHint}
\end{align}
with $G_{0}=g_{0}/\sqrt{2\pi}$, $G_{N}=g_{N}/\sqrt{2\pi}$, and $\omega_{k}=2\sqrt{t_{R}t_{L}}\cos(k)\neq E_{q}$. Here $a_{q}$ is the $q$-space annihilation operator of the lattice field, satisfying $[a_{q},a_{q'}^{\dag}]=\delta_{q,q'}$ and $[a_{q},a_{q'}]=[a_{q}^{\dag},a_{q'}^{\dag}]=0$. It is worth noting that the frequencies $\omega_{k}$ are real and in general not a function of $\beta$.

Now we use the resolvent-operator technique~\cite{resolvent}, which allows us to analytically capture the influence of a (structured) bath on quantum emitters it interacts with. For the giant emitter $c$ considered in the main text, its self-energy is given by~\cite{DLHN} 
\begin{eqnarray}
\Sigma_{c}(z)&=&\sum_{q}\frac{\mleft(G_{0}+G_{N}e^{iNq}\mright)\mleft(G_{0}+G_{N}e^{-iNq}\mright)}{z-\omega_{k}} \nonumber\\
&=&\frac{1}{2\pi}\int dq \frac{G_{0}^{2}+G_{N}^{2}+G_{0}G_{N}\mleft(e^{iNq}+e^{-iNq}\mright)}{z-2\sqrt{J_{R}J_{L}}\cos{(k)}} \nonumber\\
&=&\frac{1}{2\pi}\int dk \frac{G_{0}^{2}+G_{N}^{2}+G_{0}G_{N}e^{iNk}\mleft(\beta^{N}+\beta^{-N}\mright)}{z-2\sqrt{J_{R}J_{L}}\cos{(k)}},
\label{selfPprocess}
\end{eqnarray}
where in the last step we have changed the integration variable as $\int dq \rightarrow \int dk$ and replaced $\exp(-iNk)$ by $\exp(iNk)$ (they have identical contributions since only even functions contribute to the integral). By setting $y=\exp(ik)$ such that $2\cos(k)=y+y^{-1}$ and $\int dk=-i\oint y^{-1}dy$, and using the residue theorem, $\Sigma_{c}(z)$ can be finally obtained as
\begin{eqnarray}
\Sigma_{c}(z)&=&\frac{i}{2\pi}\oint dy\frac{G_{0}^{2}+G_{N}^{2}+G_{0}G_{N}y^{N}\mleft(\beta^{N}+\beta^{-N}\mright)}{-zy+\sqrt{J_{R}J_{L}}\mleft(y^{2}+1\mright)} \nonumber\\
&=&\mp \frac{1}{\sqrt{z^{2}-4J_{R}J_{L}}}\mleft[G_{0}^{2}+G_{N}^{2}+G_{0}G_{N}y_{\pm}^{N}\mleft(\beta^{N}+\beta^{-N}\mright)\mright],
\label{finalselfc}
\end{eqnarray}
where $y_{\pm}=\mleft(z\pm\sqrt{z^{2}-4J_{R}J_{L}}\mright)/\mleft(2\sqrt{J_{R}J_{L}}\mright)$ with the upper or lower sign chosen depending on whether $y_{+}$ or $y_{-}$ is located within the unit circle in the complex plane~\cite{AFKstructured}. The real and imaginary parts of $\Sigma_{c}(z)$ predict the frequency shift and the effective relaxation rate of emitter $c$ (induced by the lattice), respectively. For weak emitter-lattice couplings, the dynamics of the emitter can be well captured by $\Sigma_{c}(\Delta_{c}+i0^{+})$, i.e., the self-energy near the real axis~\cite{AGT2017pra,AGT2017prl}. Clearly, when $G_{N}/G_{0}=\beta^{\pm N}$, $N=2$, and $\Delta_{c}=0$ for example, $\Sigma_{c}(0+i0^{+})=0$ implying that the giant emitter is immune to relaxing into the lattice.

%%%%%%%%%%%%%%%%%%%%%%%%%%%%%%%%%%%%%%%%%%%%%%%%%%%%%%

%%%%%%%%%%%%%%%% Edited in detail by Anton up to here %%%%%%%%%%%%%%

\section{A bosonic Kitaev chain interacting with quantum emitters}
\label{appBKC}

%\begin{figure}[ptb]
%\centering
%\includegraphics[width=13 cm]{SubBKC}
%\caption{Schematic diagram of a two (small) quantum emitters coupled to a bosonic Kitaev chain (BKC). The BKC features a nearest-neighbor hopping term with amplitude (phase) $J$ ($\varphi$) and a pairing potential with strength $\Delta$. When $\varphi=\pi/2$, the position (momentum) quadratures of the two emitters are coupled to a Hatano-Nelson chain composed of the momentum (position) quadratures of the BKC site modes. The two effective Hatano-Nelson chains (i.e., the $X$ and $P$ chains) are opposite.}\label{SubBKC}
%\end{figure}

We first consider a bosonic Kitaev chain (BKC)~\cite{Aash2018prx, Aash2020nc, Chris2023arxiv}, which is described by the one-dimensional tight-binding Hamiltonian 
\begin{equation}
H_{B}=\sum_{j}\mleft(J e^{i\varphi}a_{j+1}^{\dag}a_{j}+i\Delta a_{j+1}^{\dag}a_{j}^{\dag}+\text{H.c.}\mright). \label{SMHB}
\end{equation}
Here $a_{j}$ ($a_{j}^{\dag}$) is the bosonic annihilation (creation) operator at the $j$-th lattice site; $J$ and $\varphi$ are the amplitude and phase of the nearest-neighbor hopping of the chain, respectively; $\Delta$ describes the pairing potential (i.e., two-photon drive) for every two adjacent sites. 

Now let us consider the situation where an additional bosonic mode $c$ (e.g., a quantum harmonic oscillator) is coupled to the BKC at sites $j=0$ and $j=N$, as shown in the left panel of Fig.~\figpanel{Squeeze}{a}. In this case, the total Hamiltonian of the model becomes $H=H_{B}+H_{c}+H_{\text{int}}$, with $H_{c}=\delta c^{\dag}c$ and
\begin{equation}
H_{\text{int}}=g_{0}c^{\dag}a_{0}+g_{N}c^{\dag}a_{N}+\text{H.c.}, \label{SMHint}
\end{equation}
where $\delta$ is the frequency detuning between mode $c$ and the band center of the BKC; $g_{0}$ and $g_{N}$ are the coupling strengths of $c$ to lattice sites $a_{0}$ and $a_{N}$, respectively. We refer to mode $c$ as a ``giant emitter'' if both $g_{0}$ and $g_{N}$ are nonzero and as a ``small emitter'' if $g_{N}=0$. By defining position $x_{j}=(a_{j}^{\dagger}+a_{j})/\sqrt{2}$ [$X_{c}=(c^{\dagger}+c)/\sqrt{2}$] and momentum $p_{j}=i(a_{j}^{\dagger}-a_{j})/\sqrt{2}$ [$P_{c}=i(c^{\dagger}-c)/\sqrt{2}$] quadratures for the lattice (emitter) modes, the equations of motion for the whole model are given by
\begin{align}
\dot{x}_{j}&=\mleft(J\sin{\varphi}+\Delta\mright)x_{j-1}-\mleft(J\sin{\varphi}-\Delta\mright)x_{j+1}\nonumber\\
&\quad\,+J\cos{\varphi}\mleft(p_{j+1}+p_{j-1}\mright)+\mleft(g_{0}\delta_{j,0}+g_{N}\delta_{j,N}\mright)P_{c}, \raisetag{1cm}\label{SMnewx}\\
\dot{p}_{j}&=\mleft(J\sin{\varphi}-\Delta\mright)p_{j-1}-\mleft(J\sin{\varphi}+\Delta\mright)p_{j+1}\nonumber\\
&\quad\,-J\cos{\varphi}\mleft(x_{j+1}+x_{j-1}\mright)-\mleft(g_{0}\delta_{j,0}+g_{N}\delta_{j,N}\mright)X_{c}, \raisetag{1cm}\label{SMnewp}\\
\dot{X}_{c}&=\delta P_{c}+g_{0}p_{0}+g_{N}p_{N}, \label{SMEOMxc}\\
\dot{P}_{c}&=-\delta X_{c}-g_{0}x_{0}-g_{N}x_{N}. \label{SMEOMpc}
\end{align}

The hopping phase $\varphi$ plays a crucial role in controlling the \emph{phase sensitivity} of the BKC. When $\varphi=0$ (modulo $2\pi$), Eqs.~(\ref{SMnewx}) and (\ref{SMnewp}) become
\begin{eqnarray}
\dot{x}_{j}&=&\Delta\mleft(x_{j+1}+x_{j-1}\mright)+J\mleft(p_{j+1}+p_{j-1}\mright)\nonumber\\
&&+\mleft(g_{0}\delta_{j,0}+g_{N}\delta_{j,N}\mright)P_{c}, \label{SMx0}\\
\dot{p}_{j}&=&-\Delta\mleft(p_{j+1}+p_{j-1}\mright)-J\mleft(x_{j+1}+x_{j-1}\mright)\nonumber\\
&&-\mleft(g_{0}\delta_{j,0}+g_{N}\delta_{j,N}\mright)X_{c}. \label{SMp0}
\end{eqnarray}
In this case, the quantum emitter is effectively coupled to a \emph{two-leg lattice}, where the two sublattices (we refer to them as the ``X chain'' and the ``P chain'', respectively) are coupled in an interlaced manner, i.e., with next-nearest-neighbor inter-chain couplings. The position and momentum quadratures of the quantum emitter are coupled to the P chain and the X chain, respectively, and the two emitter quadratures, $X_{c}$ and $P_{c}$, are decoupled from each other if $\delta=0$. Although such a model can be used to engineer the decay dynamics of the quantum emitter and dipole-dipole interactions between multiple emitters, it does not support phase-sensitive dynamics since the two quadratures of the lattice (i.e., the two sublattices) are always mixed in the presence of a finite pairing potential.  

On the other hand, when $\varphi=\pi/2$ (modulo $2\pi$), Eqs.~(\ref{SMnewx}) and (\ref{SMnewp}) reduce to
\begin{eqnarray}
\dot{x}_{j}&=&\mleft(J+\Delta\mright)x_{j-1}-\mleft(J-\Delta\mright)x_{j+1}\nonumber\\
&&+\mleft(g_{0}\delta_{j,0}+g_{N}\delta_{j,N}\mright)P_{c}, \label{SMxphi}\\
\dot{p}_{j}&=&\mleft(J-\Delta\mright)p_{j-1}-\mleft(J+\Delta\mright)p_{j+1}\nonumber\\
&&-\mleft(g_{0}\delta_{j,0}+g_{N}\delta_{j,N}\mright)X_{c}. \label{SMpphi}
\end{eqnarray} 
In this case, the position and momentum emitter quadratures, $X_{c}$ and $P_{c}$, are coupled to two \emph{HN chains with opposite chirality}, respectively, as shown in the right panel of Fig.~\figpanel{Squeeze}{a}. Since there are no inter-chain interactions in this case, when $\delta=0$, the model is equivalent to two distinct ``giant emitters'' (corresponding to quadratures $X_{c}$ and $P_{c}$) coupled respectively to two independent HN chains. When considering two quantum emitters, as discussed in Sec.~\ref{secTwoS} and Sec.~\ref{secGABKC}, the position (momentum) quadratures of the emitters are coupled to the P (X) chain, following the same coupling structure as the emitters themselves.

%%%%%%%%%%%%%%%%%%%%%%%%%%%%%%%%%%%%%%%%%%%%%%%%%%%%%%

\section{Self-energy for two braided emitters}
\label{appDE}

For the case of two braided emitters, it is easy to check that both emitters (i.e., $c$ and $d$, which are assumed to be identical here) do not decay into the lattice (if the non-Markovian retardation effect is neglected) by calculating their self-energies as in Sec.~\ref{appSE}. Moreover, the nonreciprocal DFI between emitters $c$ and $d$ can be explicitly understood from the two off-diagonal elements of the level-shift operator~\cite{AFKstructured,DLHN}, which are obtained as 
\begin{eqnarray}
\Sigma_{cd}(z)&=&\sum_{q}\frac{\mleft(G_{0}+G_{2}e^{2iq}\mright)\mleft(G_{1}e^{-iq}+G_{3}e^{-3iq}\mright)}{z-\omega_{k}} \nonumber\\
&=&\frac{1}{2\pi}\int dq \frac{G_{0}G_{1}\frac{e^{-ik}}{\beta}+G_{0}G_{3}\frac{e^{-3ik}}{\beta^{3}}+G_{1}G_{2}\beta e^{ik}+G_{2}G_{3}\frac{e^{-ik}}{\beta}}{z-2\sqrt{J_{R}J_{L}}\cos(k)} \nonumber\\
&=&\frac{1}{2\pi}\int dk \frac{G_{0}G_{1}\frac{e^{ik}}{\beta}-G_{0}G_{3}\frac{e^{3ik}}{\beta^{3}}-G_{1}G_{2}\beta e^{ik}+G_{2}G_{3}\frac{e^{ik}}{\beta}}{z-2\sqrt{J_{R}J_{L}}\cos(k)}, \label{SMcdprocess}\\
\Sigma_{dc}(z)&=&\sum_{q}\frac{\mleft(G_{0}+G_{2}e^{-2iq}\mright)\mleft(G_{1}e^{iq}+G_{3}e^{3iq}\mright)}{z-\omega_{k}} \nonumber\\
&=&\frac{1}{2\pi}\int dq \frac{G_{0}G_{1}\beta e^{ik}+G_{0}G_{3}\beta^{3}e^{3ik}+G_{1}G_{2}\frac{e^{-ik}}{\beta}+G_{2}G_{3}\beta e^{ik}}{z-2\sqrt{J_{R}J_{L}}\cos(k)} \nonumber\\
&=&\frac{1}{2\pi}\int dk \frac{G_{0}G_{1}\beta e^{ik}-G_{0}G_{3}\beta^{3}e^{3ik}-G_{1}G_{2}\frac{e^{ik}}{\beta}+G_{2}G_{3}\beta e^{ik}}{z-2\sqrt{J_{R}J_{L}}\cos(k)}, \label{SMdcprocess}
\end{eqnarray}
where $G_{0}=g_{0}^{2}/2\pi$, $G_{1}=\xi_{1}^{2}/2\pi$, $G_{2}=g_{2}^{2}/2\pi$, and $G_{3}=\xi_{3}^{2}/2\pi$. In the last steps of Eqs.~(\ref{SMcdprocess}) and (\ref{SMdcprocess}), we have again replaced $\exp(-ik)$ by $\exp(ik)$ like in the single-emitter case. With a similar calculation procedure, we have
\begin{eqnarray}
\Sigma_{cd}(z)&=&\mp\frac{1}{\sqrt{z-4J_{R}J_{L}}}\mleft(G_{0}G_{1}\frac{y_{\pm}}{\beta}-G_{0}G_{3}\frac{y_{\pm}^{3}}{\beta^{3}}-G_{1}G_{2}\beta y_{\pm}+G_{2}G_{3}\frac{y_{\pm}}{\beta}\mright), \label{SMselfcd}\\
\Sigma_{dc}(z)&=&\mp\frac{1}{\sqrt{z-4J_{R}J_{L}}}\mleft(G_{0}G_{1}\beta y_{\pm}-G_{0}G_{3}\beta^{3}y_{\pm}^{3}-G_{1}G_{2}\frac{y_{\pm}}{\beta}+G_{2}G_{3}\beta y_{\pm}\mright), \label{SMselfdc}
\end{eqnarray}
which, under the condition of $G_{0}=G_{1}=\beta^{2}G_{2}=\beta^{2}G_{3}$, can be simplified to
\begin{eqnarray}
\Sigma_{cd}(z)&=&\mp\frac{G_{0}^{2}}{\sqrt{z-4J_{R}J_{L}}}\mleft(\frac{2y_{\pm}}{\beta}+\frac{y_{\pm}^{3}}{\beta^{5}}+\frac{y_{\pm}}{\beta^{5}}\mright), \label{simplifiedcd}\\
\Sigma_{dc}(z)&=&\mp\frac{G_{0}^{2}}{\sqrt{z-4J_{R}J_{L}}}\mleft(\frac{2y_{\pm}}{\beta^{3}}+\beta y_{\pm}^{3}+\beta y_{\pm}\mright). \label{simplifieddc}
\end{eqnarray}
Once again, the upper or lower sign is chosen depending on whether $y_{+}$ or $y_{-}$ is located within the unit circle in the complex plane. For $z=0+i0^{+}$ (i.e., with weak emitter-lattice interactions and in the resonant case of $\Delta_{c}=\Delta_{d}=0$), we finally have 
\begin{eqnarray}
\Sigma_{P,cd}(0+i0^{+})&\simeq&\frac{-G_{0}^{2}}{J_{R}}, \label{SMfinalcd}\\
\Sigma_{P,dc}(0+i0^{+})&\simeq&\frac{-G_{0}^{2}}{J_{R}\beta^{2}}. \label{SMfinaldc}
\end{eqnarray}

The real and imaginary parts of $\Sigma_{cd}$ ($\Sigma_{dc}$) describe the \emph{coherent} and \emph{dissipative} couplings from $d$ to $c$ (from $c$ to $d$)~\cite{AFKstructured,DLHN}. Therefore, it is clear that the two braided giant emitters exhibit a nonreciprocal DFI (which is purely coherent) with the nonreciprocity opposite to that of the HN model (where the fields exhibit rightward amplification).

\end{widetext}

\bibliography{GAskin_ref}

\end{document}